\documentclass[prb,aps,epsf,twocolumn,superscriptaddress]{revtex4-1}
\usepackage{graphicx,amsfonts,times,bm,amsmath,verbatim,color,array}

\newcommand{\ket}[1]{| {#1} \rangle}

\usepackage{subfigure}
\usepackage{float}
\usepackage{epsfig}
\usepackage{hyperref}
\hypersetup{
colorlinks=true,final=true,
        linkcolor=blue,
        citecolor=blue,
        filecolor=blue,
        urlcolor=blue,
}

\begin{document}
\title{Mirror anomaly and anomalous Hall effect in type-I Dirac semimetals}

\author{S. Nandy}
\affiliation{Max-Planck Institute for the Physics of Complex Systems, D-01187 Dresden, Germany}
\affiliation{Department of Physics, Indian Institute of Technology Kharagpur, W.B. 721302, India}
\author{Kush Saha}
\affiliation{Max-Planck Institute for the Physics of Complex Systems, D-01187 Dresden, Germany}
\affiliation{School of Physical Sciences, National Institute of Science Education and Research,  Jatni 752050, Khurda, India}
\author{A. Taraphder}
\affiliation{Department of Physics, Indian Institute of Technology Kharagpur, W.B. 721302, India}
\affiliation{Centre for Theoretical Studies, Indian Institute of Technology Kharagpur, W.B. 721302, India}
\affiliation{School of Basic Sciences, Indian Institute of Technology Mandi, Kamand 175005, India}
\author{Sumanta Tewari}
\affiliation{Department of Physics and Astronomy, Clemson University, Clemson, SC 29634,U.S.A}


\begin{abstract}
In addition to the well known chiral anomaly, Dirac semimetals have been argued to exhibit mirror anomaly, close analogue to the parity anomaly of  ($2+1$)-dimensional massive Dirac fermions. The observable response of such anomaly is manifested in a singular step-like anomalous Hall response across the mirror-symmetric plane in the presence of a magnetic field. Although this result seems to be valid in type-II Dirac semimetals (strictly speaking, in the linearized theory), we find that type-I Dirac semimetals do not possess such an anomaly in anomalous Hall response even at the level of the linearized theory. In particular, we show that the anomalous Hall response continuously approaches zero as one approaches the mirror symmetric angle in a type-I Dirac semimetal as opposed to the singular Hall response in a type-II Dirac semimetal. Moreover, we show that, under certain condition, the anomalous Hall response may vanish in a linearized type-I Dirac semimetal, even in the presence of time reversal symmetry breaking. 
\end{abstract}

\maketitle

\section{Introduction}
Nearing more than a decade of rapid progress in the field of two-dimensional Graphene~\cite{Neto_2009,Sarma_2011}, the theoretical predictions and discoveries of its 3D analogue, namely Dirac semimetals (DSMs) have led to an explosion of activity in recent  years~\cite{Kane_2010,Wan:2011,Yang_2011, Neupane_2014, Xu_2011, Xu_2015, Balents_2011, Young_2012, Liu_2014, Wang_2012, Wang_2013}.
Due to their distinct linearly dispersing bulk bands, 3D Dirac semimetals are found to exhibit several unconventional electronic properties such as giant diamagnetism~\cite{Wang_2012}, oscillating quantum spin Hall effect~\cite{Wang_2012, Burkov_2016}, quantum magnetoresistance\cite{He:2014, Li_2013, Xiong}, and many more\cite{Gorbar_2018,Lin_2017}. Moreover, they are known to host exotic quantum states such as topological insulators and Weyl metals under some external perturbations and symmetries such as time-reversal symmetry (TRS), inversion symmetry (IS) and crystalline uniaxial rotational symmetry $C_{n}$. With these symmetries, DSMs can be divided into two classes- type-I and type-II\cite{Young_2012, Young_2014, Nagaosa_2014}. In type-I DSMs, Dirac points generally occur in pairs on the rotation axis~\cite{Liu_2014, Wang_2013} but away from the time reversal invariant momenta (TRIM). The Dirac points here can be obtained through an accidental band crossing or a band inversion. However, they can be stabilized by crystalline symmetry other than the TRS and IS. In contrast, in type-II DSMs, a single Dirac point occurs at the TRIM on the rotation axis\cite{Young_2012, Mele_2018}, and band crossing is ensured by the lattice symmetry. A series of recent experiments confirmed that Na$_{3}$Bi~\cite{Liu_2014, Jeng_2015, Krizan_2015} and Cd$_{3}$As$_{2}$~\cite{He:2014, Liang:2015, Neupane_2014, Mo_2014, Jeon_2014} compounds show type-I behavior, while TlBi(S$_{1-x}$Se$_x$)$_2$~\cite{Sato_2011}, (Bi$_{1-x}$In$_x$)$_2$Se$_3$~\cite{Brahlek_2012}, and ZrTe$_5$~\cite{Chen_2015, Valla_2016} belong to the type-II Dirac semimetals.

The distinction of DSMs may not be limited to the number and location of Dirac points in the Brillouin zone, rather it may be manifested in the observable responses in terms of quantum anomaly in the presence of an external magnetic field. The principal among such quantum anomalies is the chiral anomaly which indicates the non-conservation of electric charge of a given chirality in the presence of  parallel electric and magnetic fields. Although both types of Dirac semimetals are supposed to show negative longitudinal magnetoresistance and planar Hall effect as a manifestation of chiral anomaly in the presence of parallel electric and magnetic field~\cite{Bell:1969, Adler:1969, Nielsen:1981, Nielsen:1983, Goswami:2015, Zhong, Aji:2012, Goswami:2013, Zyuzin:2012, Nandy1_2017, Nandy_2017}, there may exist another anomaly in DSMs namely mirror anomaly which may give rise to different observable signatures in type I and type II DSMs in the anomalous Hall effect. In particular, this anomaly gives rise to step-function dependence of the anomalous Hall conductivity (AHC) as a function of the polar angle of the applied magnetic field, resembling the AHC due to parity anomaly in 2D systems with massive Dirac fermions~\cite{Haldane_1988}. Recently it has been argued that type-II DSMs show a singular step-like Hall response across the mirror symmetric angle of an applied magnetic field due to the mirror anomaly~\cite{Burkov_2017}. This raises interesting questions: Is the Hall response due to the mirror anomaly a generic feature of Dirac semimetals? In other words, do type-I Dirac semimetals also possess such anomaly?

To answer these, we consider a model Hamiltonian of a type-I DSM, in particular Hamiltonian of  Cd$_{3}$As$_{2}$, and study the anomalous Hall effect in the presence of a rotating magnetic field. It is important to note that neither type-I nor type-II DSMs are expected to show an anomalous Hall effect in the absence of a magnetic field because of the presence of time reversal symmetry. In the presence of a Zeeman field, however, time reversal is explicitly broken, resulting in the creation of pairs of Weyl points with charges of opposite chirality, which may produce anomalous Hall effect. Following Ref.~\onlinecite{Burkov_2017}, by anomalous Hall effect in DSMs we consider the component of the total Hall response created by the Zeeman effect of the applied magnetic field, separable from the conventional, orbital, effect by  its linear $B$ dependence. We find that type-I DSMs  do not possess mirror anomaly across the plane preserving mirror symmetry in the presence of an external magnetic field. Specifically, we do not find any step-function like behavior in the AHC as a function of the angle of the applied field. This is in sharp contrast to the type-II DSMs\cite{Burkov_2017}. We furthermore show that, within linearized theory, the AHC vanishes under certain conditions. Thus AHC may be used as a probe to identify two distinct types of Dirac semimetals.

The rest of the paper is organized as follows. In Sec.~\ref{Model Hamiltonian}, we introduce the low energy model Hamiltonian of a time-reversal- and inversion symmetric type-I Dirac semimetal. We then discuss possible mirror symmetric planes and relevant mirror-reflection operators. This is followed by the review of the anomalous Hall response in a type-II DSM in Sec.~\ref{type_II_hall}, where we compute Hall conductivity in the presence of a magnetic field applied perpendicular to the mirror symmetric plane and discuss the possibility of appearing mirror anomaly. In Sec.~\ref{type-I_hall}, we compute AHC in a type-I DSM and compare the results with the type-II DSM. Finally, we summarize our results and discuss possible future direction in Sec.~\ref{summary}.

\section{Model Hamiltonian}
\label{Model Hamiltonian}

\begin{figure}[htb]
	\begin{center}
		\epsfig{file=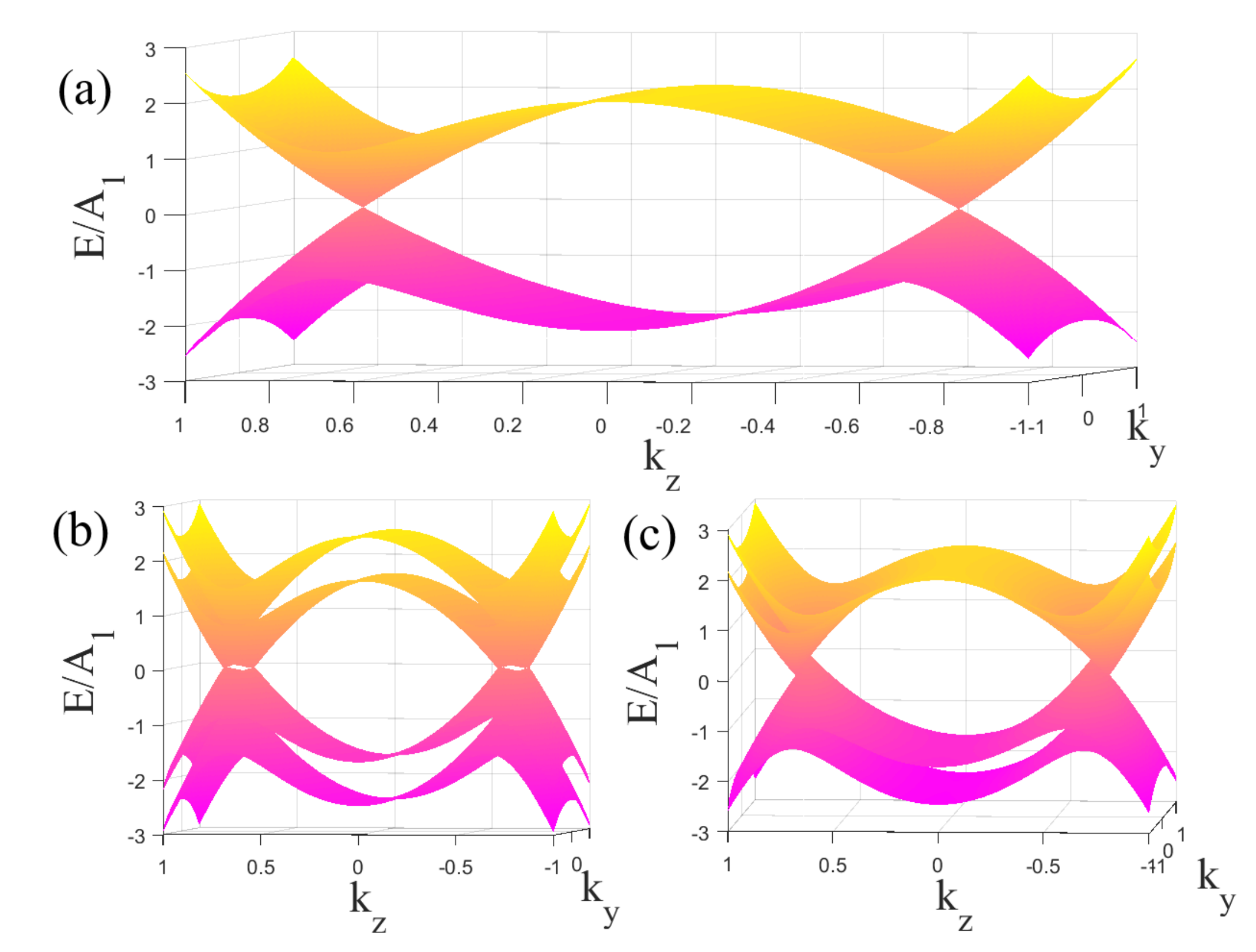,trim=0.0in 0.05in 0.0in 0.05in,clip=true, width=90mm}\vspace{0em}
		\caption{(Color online) (a) 3D band dispersion of the four bands ($k_{x}$ is suppressed) of the Dirac semimetal near $\Gamma$ point obtained by diagonalizing Hamiltonian described in Eq.~(\ref{H_main}). The Dirac points are located at $(0,0,\pm k_{z}^0)$ where $k_{z}^0=\sqrt{\frac{M_{0}}{M_{z}}}$. (b)-(c) depict the band dispersions in the presence of magnetic field (B) applied along $z$ direction and $x$ direction respectively. The parameters are chosen to be $M_{0}=-2A_{1}$, $M_{\parallel}=-A_{1}/5$, $M_{z}=-4A_{1}$, $\beta=-A_{1}/5$ and $\gamma=A_{1}$, $B=A_{1}/5$ and $A_{1}=0.05 eV$.~\cite{Sato_2015}}
		\label{disperse}
	\end{center}
\end{figure}

We begin with a discussion on the model Hamiltonian of type-I Dirac semimetals, containing a pair of four-fold degenerate Dirac nodes on a high symmetry axis. The Dirac semimetals Na$_{3}$Bi and Cd$_{3}$As$_{2}$, are both thought to be in this class. In both of these materials, there are two pairs of relevant orbitals near the Fermi level at the $\Gamma$ point : the s orbitals with $j_{z} = \frac{1}{2}$ and
the p orbitals with $j_{z} = \frac{3}{2}$. Because of the atomic spin-orbit coupling, band inversion occurs in both cases and the band crossing occurs at the Fermi level near the $\Gamma$ point. The pair of Dirac points occur along the $k_{z}$ axis and stabilized due to discrete rotation symmetry about $k_{z}$ axis ($C_{4z}$ for Cd$_{3}$As$_{2}$ and $C_{6z}$ for Na$_{3}$Bi). In the basis of the relevant spin-orbit
coupled states $\ket{s,\uparrow}$, $\ket{p_{x}+ip_{y},\uparrow}$, $\ket{s,\downarrow}$ and $\ket{p_{x}-ip_{y},\downarrow}$, the low energy effective $\mathbf{k} \cdot \mathbf{p}$ Hamiltonian for the type-I Dirac semimetals near the $\Gamma$ point can be written as~\cite{Nagaosa_2014, Hashimoto_2016, Bernevig_2017,  Chen_2016}

\begin{align}
H^{\rm I}(\mathbf{k})&= M(\mathbf{k})\tau_{z}\sigma_{0}+ A(\mathbf{k})\tau_{x}\sigma_{z}+ B(\mathbf{k})\tau_{y}\sigma_{0}+ C(\mathbf{k})\tau_{x}\sigma_{x} \nonumber \\
&~~~~~~~~~~~~~~~~~~~~~~~~~~~~~~~~~~~~~~~~~~~~~~~~~~~+ D(\mathbf{k})\tau_{x}\sigma_{y}\label{H_main},
\end{align}
where $\tau^i$ and $\sigma^i$ are Pauli matrices in orbital and spin space respectively, $\quad k_{\parallel}=\sqrt{k_{x}^2+k_{y}^2}$ is the momentum parallel to the surface of the DSM, $M({\mathbf{k}})=M_{0}-M_{\parallel}k_{\parallel}^{2}-M_{z}k_{z}^{2}$ is the Schrodinger/Dirac mass term, $A({\mathbf{k}})=A_{1}k_{x}$, $\quad \quad B({\mathbf{k}})=-A_{1}k_{y}$,
$C({\mathbf{k}})=(\beta+\gamma)k_{z}(k_{y}^{2}-k_{x}^{2})$, and $\quad D({\mathbf{k}})=-2(\beta-\gamma)k_{z} k_{y} k_{x}$.
The band parameters M$_{0}$, M$_{\parallel}$, M$_{z}$, A$_{1}$, $\beta$ and $\gamma$ are material dependents and can be obtained from the first principles calculations. The low-energy spectrum of Eq.~(\ref{H_main}) contains a pair of Dirac points located at $(0,0,\pm k_{z}^0)$ where $k_{z}^0=\sqrt{\frac{M_{0}}{M_{z}}}$ as shown in Fig.~(\ref{disperse}a).

Eq.~(\ref{H_main}) is invariant under time-reversal $({\mathcal T}=i\tau_{0}\sigma_{y}\it{K})$ and inversion symmetry $(I=\tau_{z}\sigma_{0})$, where $K$ is complex conjugation operator\cite{Hashimoto_2016}. Additionally, Eq.~(\ref{H_main}) has several mirror symmetric planes such as $(100)$, $(001)$ and $(110)$~\cite{Hashimoto_2016}. For simplicity, we focus only on $(100)$ plane i.e. $yz$ plane. Our results will be qualitatively same for other mirror symmetric planes. In the $yz$ plane, the mirror symmetry operator is given by $M_{yz}=i\tau_0\sigma_x$, satisfying
\begin{align}
M_{yz}H(k_x,k_y,k_z)M^{\dagger}_{yz}=H(-k_x,k_y,k_z)
\end{align}
Note that the mirror symmetry appears here as a result of the combination of space inversion symmetry and a rotation about the $x-$ axis. This follows from the existence of the $C_{n}$ symmetry~\cite{Nagaosa_2014}.

\section{Anomalous Hall Conductivity}
\label{Anomalous_Hall}

Let us now consider an external magnetic field applied to the Dirac semimetal. The presence of magnetic field breaks time-reversal symmetry and leads to non-zero Berry curvature. Consequently, one may obtain  anomalous Hall conductivity induced by the non-trivial Berry curvature.
The intrinsic contribution of the anomalous Hall effect of a three dimensional system can be written as~\cite{Xiao_2010}
\begin{eqnarray}
\sigma_{ij}=\frac{e^{2}}{\hbar}\frac{1}{(2\pi)^3} \int d^3{\bf k} \Omega_{k_i,k_j} f_{0},
\label{e2}
\end{eqnarray}
where $f_{0}$ is the equilibrium Fermi-Dirac distribution function and $\Omega_{k_i,k_j}$ is the Berry curvature in the direction perpendicular to $k_i-k_j$ plane. The Berry curvature of the $n^{th}$ band for a Bloch Hamiltonian $H({\bf k})$, defined as the Berry phase per unit area in the $k$ space, is given by~\cite{Xiao_2010}
\begin{eqnarray}
\Omega_{k_i,k_j}=2i\sum_{n \neq n^{\prime}} \frac{\langle n|\frac{\partial H}{\partial k_{i}}|n^{\prime}\rangle\langle n^{\prime}|\frac{\partial H}{\partial k_{j}}|n\rangle}{(\epsilon_{n}-\epsilon_{n^{\prime}})^{2}},
\label{Berry}
\end{eqnarray}
where $\epsilon_{n}$ is the energy dispersion on $n-$th band and $|n\rangle$ is the corresponding eigenstate.

\subsection{Type-II Dirac semimetals}
\label{type_II_hall}

To compare anomalous Hall response of a type-I DSM with type-II DSM, we first review the response in the type-II DSM as discussed in Ref.~\onlinecite{Burkov_2017}. The linearized low-energy model Hamiltonian of a type-II Dirac semimetal is given by\cite{Burkov_2017}
\begin{eqnarray}
H^{\rm II}(k)=v(-k_x \tau_{z}\sigma_{y}+k_y \tau_{z}\sigma_{x}+k_z \tau_{y}\sigma_{0}),
\label{type-II}
\end{eqnarray}
where $v$ is the quasiparticle velocity and $\tau$'s and $\sigma$'s are orbital and spin degrees of freedom of the corresponding model. Eq.~(\ref{type-II}) contains single Dirac node at $\Gamma$ point. In addition, it  possesses mirror symmetry in the $yz$ plane and $xz$ plane as follows:

\begin{figure}[t]
	\begin{center}
		\epsfig{file=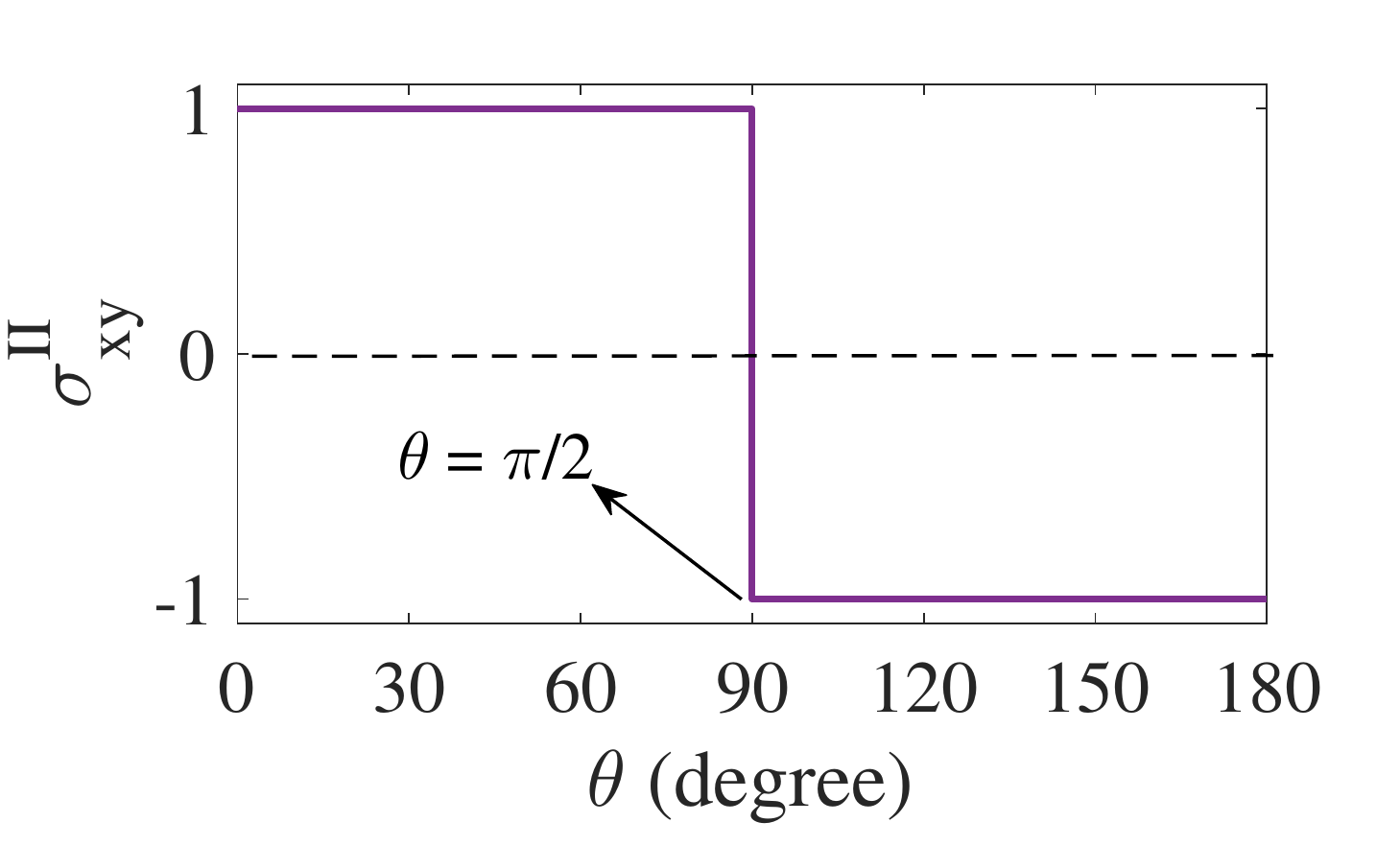,trim=0.0in 0.05in 0.0in 0.05in,clip=true, width=90mm}\vspace{0em}
		\caption{(Color online)  Normalized anomalous Hall conductivity of a type-II Dirac semimetal as a function of $\theta$. Here, $\theta$ is the angle between $z$ axis and applied magnetic field. The sign of AHC changes at the mirror symmetric angle  ($\theta=\pi/2$).}
		\label{AH2}
	\end{center}
\end{figure}

\begin{align}
M_{yz}H^{\rm II}(k_x,k_y,k_z)M_{yz}^{\dagger}=H^{\rm II}(-k_x,k_y,k_z)\nonumber\\
M_{xz}H^{\rm II}(k_x,k_y,k_z)M_{xz}^{\dagger}=H^{\rm II}(k_x,-k_y,k_z),
\end{align}
where $M_{yz}=i\sigma_x$ and $M_{xz}=i\sigma_y$. These mirror-reflection symmetries have observable consequences in the presence of a magnetic field as will be evident shortly.

In what follows, we consider a magnetic field in the $xz$ plane as ${\bf B}=B(\sin\theta,0, \cos\theta)$ where $\theta$ is the angle between the $z$ axis and $B$. This leads to the Zeeman term
\begin{align}
H_{\rm zmn}^{\rm II}(\theta)= b\cos\theta\sigma_z+b\sin\theta\sigma_x,
\label{zeeman}
\end{align}
where $b=g\mu_{B}B$, $g$ is the Lande g factor and $\mu_B$ is the Bohr magneton. For simplicity, we ignore orbital coupling of the magnetic field to the Dirac electrons. In the presence of $H_{\rm zmn}^{\rm II}$, the single Dirac node splits into two Weyl nodes with opposite chirality located at $k_z=\pm b/v$. For a fixed $B$, the location of these Weyl nodes does not move as we vary $\theta$ as clearly seen from Fig.~4. However, the Weyl nodes interchange their topological charge with each other as $\theta$ crosses mirror symmetric value, i.e., $\theta=\pi/2$. This fact will be manifested in the Hall conductivity.
Diagonalizing $H^{\rm II}(k)+H_{\rm zmn}^{\rm II}(\theta)$ numerically, it is straight forward to obtain Hall conductance $\sigma_{xy}$ as a function of $\theta$. However, it does not provide an intuitive understanding of the anomalous behavior of the conductance related to the mirror anomaly. Thus it is desirable to have analytical form of the Hall conductivity.
In doing so, we block-diagonalize $H^{\rm II}(k)+H_{\rm zmn}^{\rm II}(\theta)$ by rotating the spin quantization axis along the direction of ${\bf B}$ and using similarity transformations\cite{Burkov_2017} $\sigma^{x,y}\rightarrow \tau^z\sigma^{x,y}$ and $\tau^{x,y}\rightarrow \sigma^z\tau^{x,y}$, we obtain
\begin{align}
H_{2\times 2}=v k_y\cos\theta\sigma_x-vk_x\sigma_y+m_r\sigma_z,
\end{align}
where $m_r=b+rv\sqrt{k_y^2\sin^2\theta+k_z^2}$ with $r=\pm$. With this, the $z$-component of Berry curvatures of lowest two bands read off
\begin{align}
\Omega_z({\bf k})^{\pm}= \frac{v^2\cos\theta(b\pm \frac{vk_z^2}{\sqrt{k^2_y\sin^2\theta+k_z^2}})}{4\epsilon_{k, \pm}^{3}},
\end{align}
where $\epsilon_{k,\pm}=\sqrt{v^2k_x^2+v^2k_y^2\cos^2\theta+m_{\pm}^2}$. For Fermi energy in the middle of the gap, i.e., $\epsilon_F=0$, the contribution from both the Berry curvatures $\Omega_z({\bf k})^+$ and $\Omega_z({\bf k})^-$ lead to
\begin{align}
\sigma_{xy}(\theta)=\frac{e^2}{h}\frac{b}{v\pi} sgn(\cos \theta)
\label{AH-I}
\end{align}
Evidently at $\theta=\pi/2$, the $\sigma_{xy}$ vanishes due to mirror symmetry. However, for $\theta\rightarrow \pi/2$, it does not
vanish from both side of $\theta=\pi/2$. This fact is called mirror anomaly analogous to the parity anomaly\cite{Haldane_1988, Semenoff_1984} in a massive ($m$) Dirac Fermion with Hall conductivity $\sigma_{xy}=\frac{e^2}{2h}sgn(m)$. Thus we show that the anomalous Hall conductivity described in  Eq.~(\ref{AH-I}) indicates mirror anomaly in the linearized type-II DSMs. The AHC as a function of $\theta$ for type-II DSMs is depicted in Fig.~\ref{AH2}. The sign change of AHC at $\theta=\pi/2$ can be attributed to the interchange of topological charge of the Weyl nodes with each other at the mirror symmetric plane. It is worth pointing out that the addition of cubic term $(k_z^2-k_x^2)k_y\tau_z\sigma_z$ in Eq.~(\ref{type-II}) breaks mirror symmetry, hence the absence of step-like function across the mirror symmetric plane can be obtained.

\begin{figure}[t]
	\begin{center}
		\epsfig{file=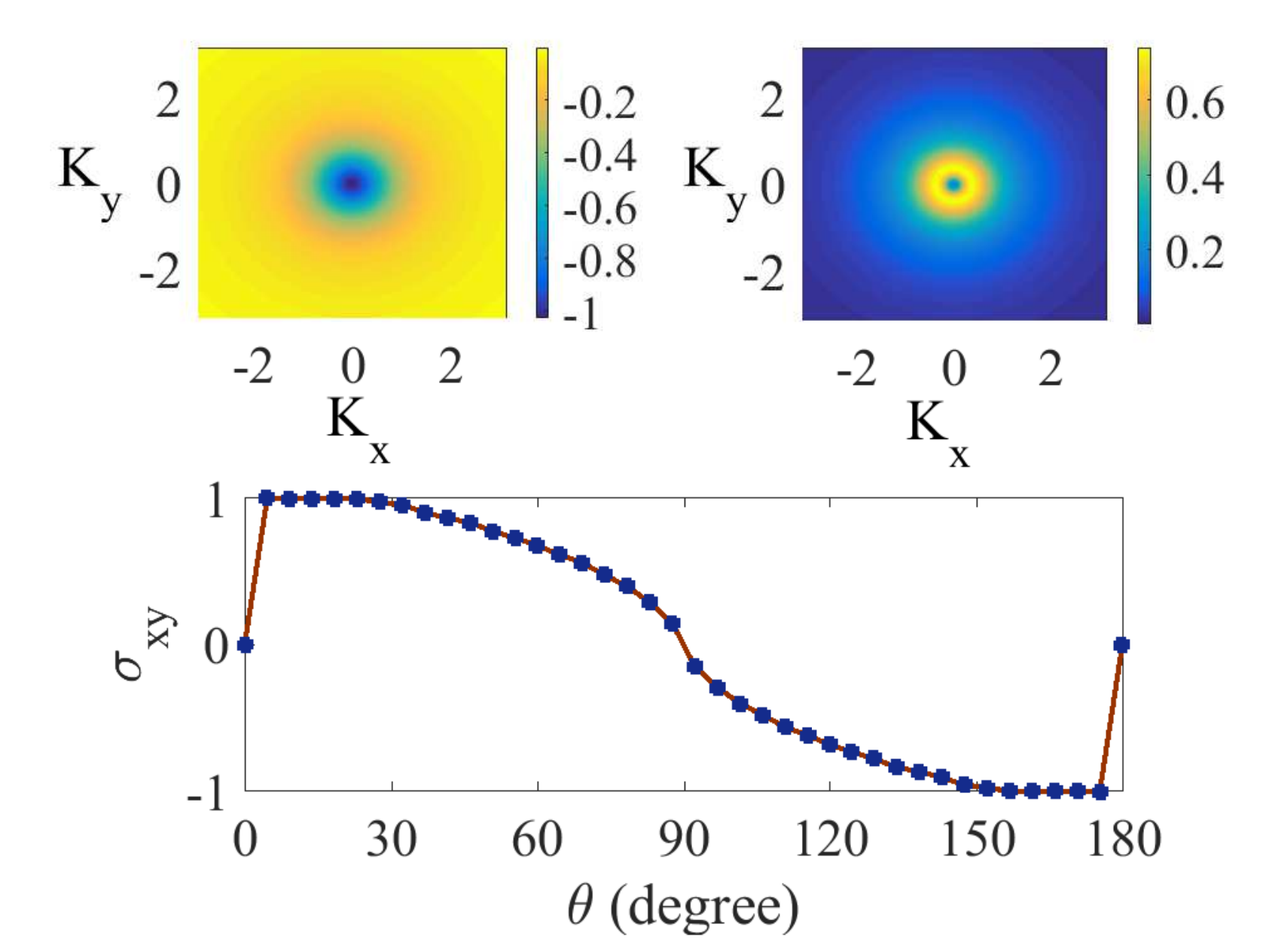,trim=0.0in 0.05in 0.0in 0.05in,clip=true, width=90mm}\vspace{0em}
		\caption{(Color online) (a)-(b) depict $z$ component of the Berry curvature ($\Omega_{z} (\mathbf{k})$) at $\theta=\pi/4$ in $k_x-k_y$ plane for lowest two bands of type-I Dirac semimetal respectively. Here we have neglected the cubic terms of the Hamiltonian described in Eq.~(\ref{H_main}). (c) depicts the normalized AHC for type-I DSM in the absence of cubic terms.}
		\label{Bery_cur}
	\end{center}
\end{figure}

\subsection{Type-I Dirac semimetals}
\label{type-I_hall}
Having discussed the Hall response in a type-II DSM, we now investigate if similar physics holds in a  type-I DSM. To study this, we consider a magnetic field rotated in the $xz$ plane as before. The Zeeman field term can be written as~\cite{Hashimoto_2016}
\begin{align}
H_{\rm zmn}^{\rm I}(\theta)=b\sin\theta\frac{(\tau_{0}+\tau_{z})}{2}\sigma_{x}+b\cos\theta\tau_{0}\sigma_{z}.
\label{type_I_zeeman}
\end{align}



To compare with the results obtained from the linearized form of the type-II Hamiltonian in Eq.~(\ref{type-II}), we first neglect the cubic terms $C(\mathbf{k})$ and $D(\mathbf{k})$ in Eq.~(\ref{H_main}). Note that, in this case, a pair of momentum-resolved energy degenerate Dirac points occur at $(0,0,\pm k_z^{0})$. The presence of magnetic field in the $z$-direction $(\theta=0)$ lifts the degeneracy of the four bands at each Dirac point and leads to a double pair of Weyl points along the same axis as shown in Fig.~\ref{disperse}b. These four Weyl points are located at  $(0,0,\pm \sqrt{\frac{\pm B_{z}+M_{0}}{M_{z}}})$. However, for $\theta\ne 0$, the rotated magnetic field gives rise to momentum resolved Weyl points that move in the energy momentum phase space as $\theta$ is varied. 
First, we compute the Hall conductance of the Hamiltonian in Eq.~(\ref{H_main}) (with $C=D=0$) when the magnetic field is along the $z$-direction. This particular case allows us to write $H_{I}+H^I_{zmn}(\theta=0)$ in the block-diagonalized form as
\begin{eqnarray}
{\mathcal H}_r=[M({\bf k})\sigma_z-A_1 k_y\sigma_y+r(A_1 k_x\sigma_x+b\sigma_0)],
\end{eqnarray}
where $r=\pm$ are the two eigenvalues of $\tau_z$. Then the Berry curvatures of the lowest two bands turn out to be
\begin{align}
\Omega({\bf k})=\mp \frac{(M_{0}+M_{\parallel}k_{\parallel}^{2}-M_{z}k_{z}^{2})A_1^2}{4E({\bf k})^{3}},
\end{align}
where $E({\bf k})=\sqrt{M({\bf k})^2+A_1^2(k_x^2+k_y^2)}$. Evidently, $\Omega$ for the lowest two bands are equal and opposite in sign. Thus for the Fermi level ($\epsilon_F$) in the middle of the band spectrum, the sum of the two Berry curvatures vanishes, and so does the AHC.
This is in contrast to the case of type-II DSM, where the magnetic field along $z$ gives finite conductivity (see Fig.~\ref{AH2}).
\begin{figure}[htb]
\centering
\begin{tabular}{cc}
\includegraphics[width=44mm]{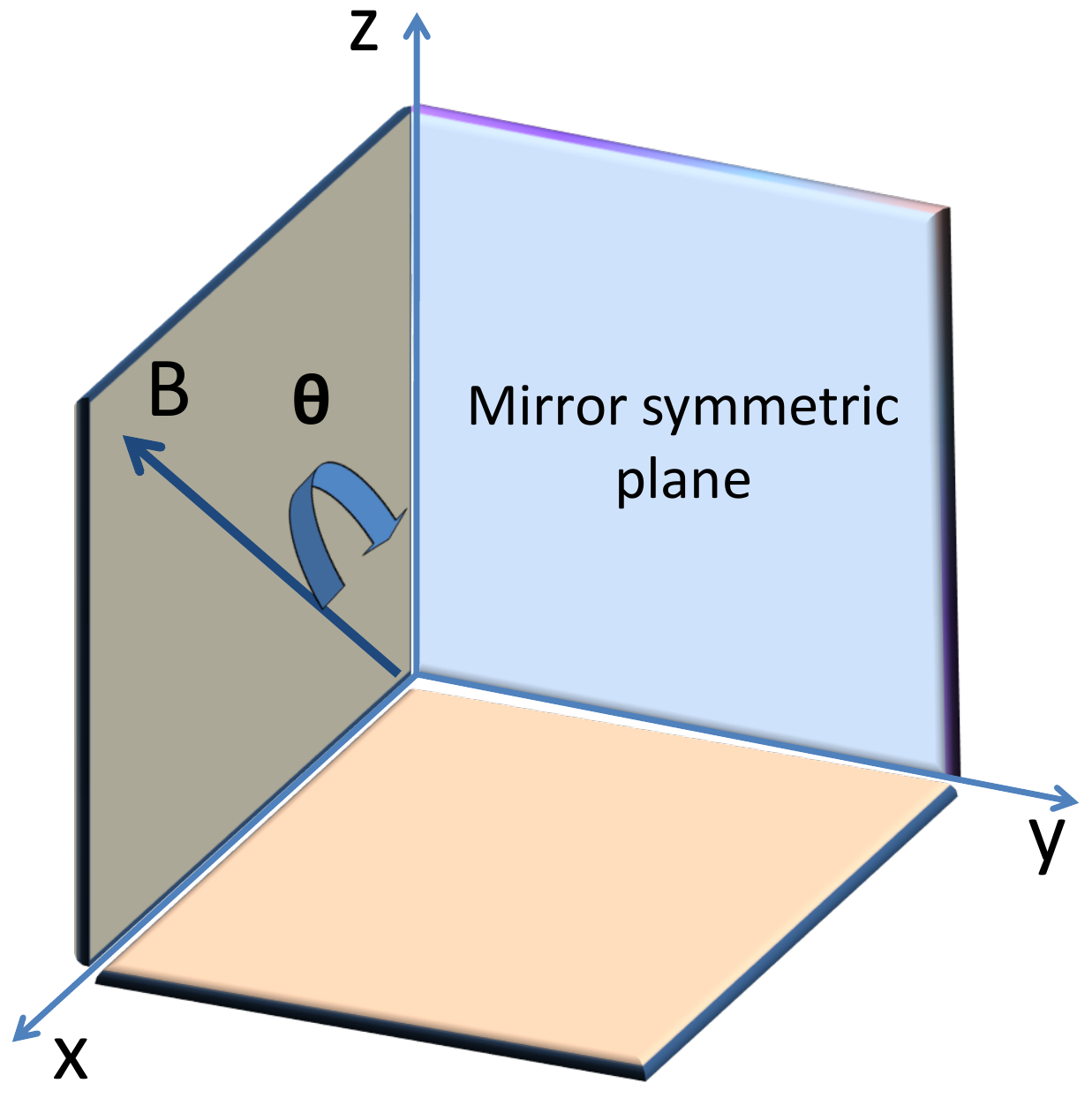}
\includegraphics[width=45mm]{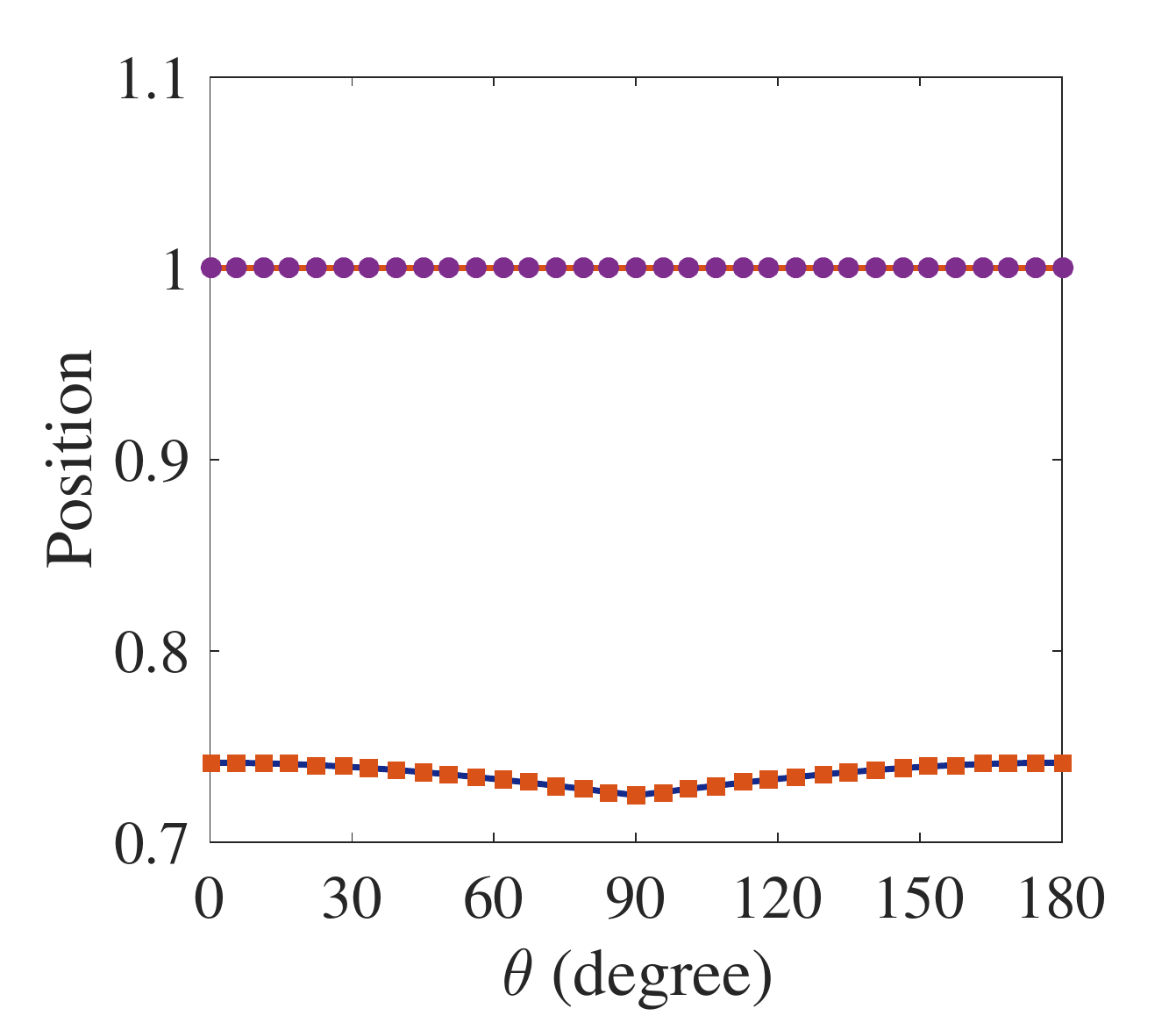}
\end{tabular}
\label{P_W}
\caption{(Color online) (Left) Schematic diagram of the mirror symmetric plane ($yz$) and the applied magnetic field is rotated in $xz$ plane perpendicular to the mirror symmetric plane. (Right) depicts the position of the rightmost Weyl node as a function of $\theta$ for type-II (Upper curve) and type-I (Lower curve) Dirac semimetals. It is clear that the position of Weyl node is changing with the rotation of magnetic field in type-I DSM whereas the position of Weyl node remains same in type-II DSM. The values of the parameters are same as mentioned in caption of Fig~1.}
\end{figure}
We next move to the case of rotated magnetic field ($\theta\ne0$), but still within the linearized theory. This case does not allow us to express $H_{I}+H^I_{zmn}(\theta)$ in a block-diagonal form. Thus we compute Hall conductance numerically using Eq.~(\ref{Berry}). The total contribution to the AHC in type-I DSM comes from the region between the Weyl nodes. Since in our case, type-I DSM possesses four Weyl nodes in the presence of B, there exist three such regions. Fig.~\ref{Bery_cur}(a)-(b) illustrate $\Omega ({\bf k})$ of the lowest two bands for arbitrary $\theta$ ($\theta=\pi/4$) for a particular $k_z$ value. It turns out that the  $\Omega ({\bf k})$'s are opposite in sign but unequal in magnitude as opposed to the case of $\theta=0$. Thus, the sum of the Berry curvatures gives rise to finite contribution to the AHC as shown in Fig.~(\ref{Bery_cur}c). It is clear from the figure that the magnitude of AHC changes smoothly across the mirror symmetric angle $\theta=\pi/2$.
We furthermore checked that, for magnetic field in the $xy$ plane, $\Omega$'s for each band identically vanishes. Thus, indeed, the low-energy linearized Hamiltonian of type-I DSM does not show any signature of mirror anomaly and this is in contrast to the type-II DSM where the linearized Hamiltonian produces a sharp step-function like signature in the AHC at the mirror symmetric angle $\theta=\pi/2$.

\begin{figure}[htb]
	\begin{center}
		\epsfig{file=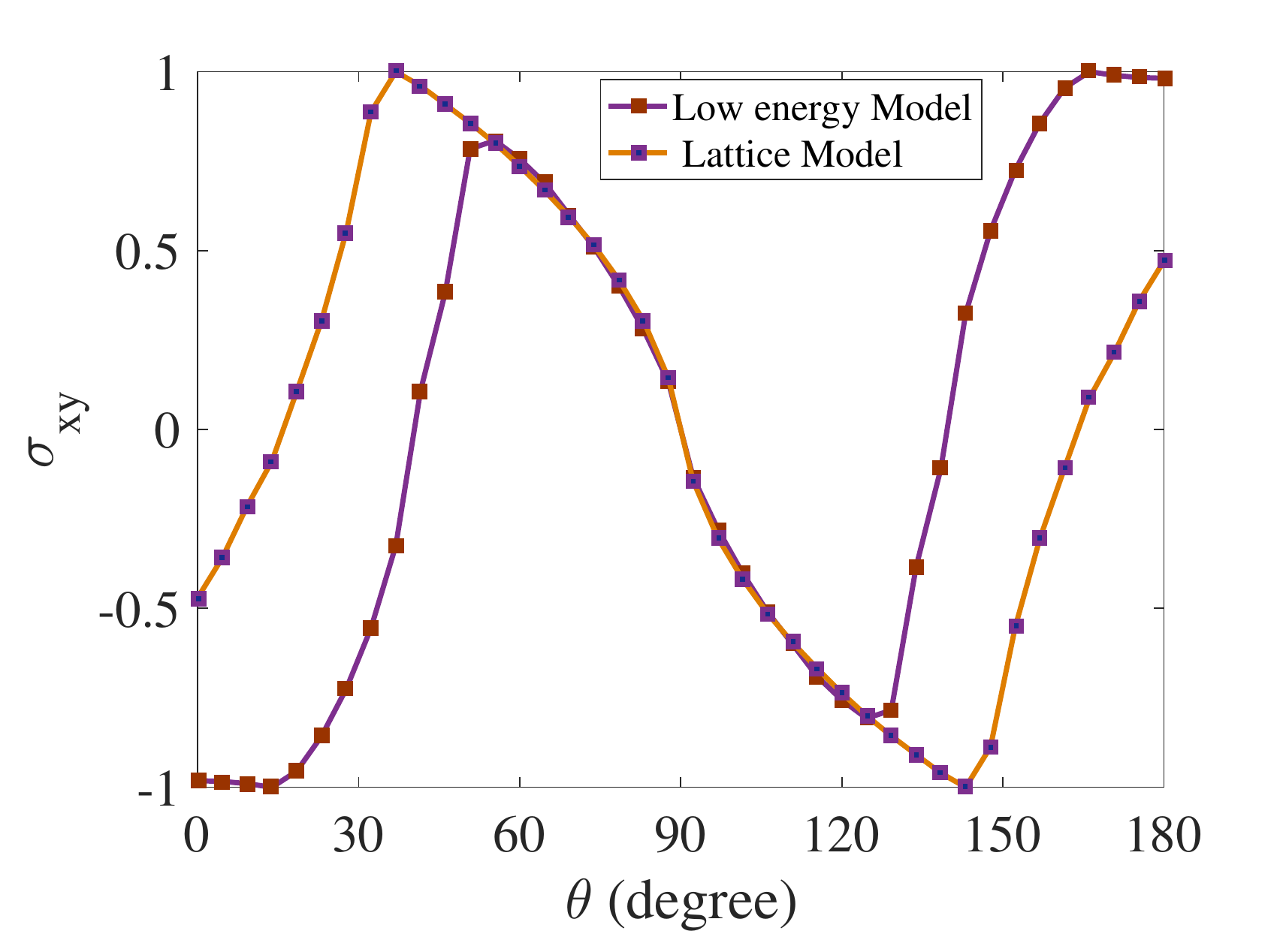,trim=0.0in 0.05in 0.0in 0.05in,clip=true, width=90mm}\vspace{0em}
		\caption{(Color online) Normalized anomalous Hall conductivity of type-I Dirac semimetal as a function of $\theta$ for both the low energy Hamiltonian (Eq.~(\ref{H_main})) and the full lattice model (Eq.~(\ref{H_Lattice})). Here, $\theta$ is the angle between $z$ axis and the applied magnetic field rotated in $xz$ plane, see Fig 4. The sign of AHC changes at the mirror symmetric angle ($\theta=\pi/2$) while the magnitude evolves smoothly. The results of the low energy Hamiltonian and the full lattice model are qualitatively consistent, as is expected for anomalous Hall effect. The values of the parameters of the low energy model of Eq.~(\ref{H_main}) are given in caption of Fig.~1. The values of the parameters for the lattice model in Eq.~(\ref{H_Lattice}) are taken so as to be consistent with the low energy model of Eq.~(\ref{H_main}).}
		\label{AH1}
	\end{center}
\end{figure}

Including the cubic terms $C({\bf k})$ and $D({\bf k})$ in Eq.~(\ref{H_main}) does not allow simple block-diagonal form for the Hamiltonian even for the magnetic field along the $z$-direction. Thus we again numerically find eigenstates of $H^{\rm I}+H_{\rm zmn}^{\rm I}(\theta)$ and use them in Eq.~(\ref{Berry}) to obtain AHC. Fig.~\ref{AH1} shows the anomalous Hall conductivity for the full Hamiltonian in Eq.~(\ref{H_main}) as a function of $\theta$.
It is clear from the figure that magnitude of AHC changes smoothly as the angle $\theta$ approaches $\pi/2$ and becomes zero at $\theta=\pi/2$ as a manifestation of mirror symmetry. After passing through the mirror symmetric angle, AHC changes sign from positive to negative and its magnitude increases smoothly. Thus, in contrast to type-II DSM, in which the behavior of AHC as a function of the polar angle $\theta$ of the applied magnetic field is akin to a broadened step-function centered at the mirror symmetric value $\theta=\pi/2$, in type-I DSM the AHC in the presence of the cubic terms of the Hamiltonian, smoothly evolves as a function of the polar angle, without showing any evidence of mirror anomaly. In order to get a deeper understanding of our result we plot the position of a representative Weyl node (the Weyl node located at (0, 0, $\sqrt{\frac{- B_{z}+M_{0}}{M_{z}}}$) for $\theta=0$) of Eq.~(\ref{H_main}) in the presence of a magnetic field as a function of $\theta$ as shown in Fig.~4. Evidently, the positions of the Weyl points change with the angle ($\theta$) smoothly in such a way that the separation between the opposite chirality Weyl nodes also changes smoothly. That is why the magnitude of AHC varies smoothly across the mirror symmetric value of the polar angle $\theta$ of the applied magnetic field.


So far, we have used the low energy Hamiltonian given in Eq.~(\ref{H_main}) to calculate the AHC in the presence of a rotating external magnetic field. The low energy Hamiltonian is useful because it allows, in some cases, to analytically evaluate the Berry curvatures, and also to compare the results with the low energy treatment of type-II DSM given in Ref.~\onlinecite{Burkov_2017}. To confirm our main conclusion -- the absence of mirror anomaly signature in AHC in type-I DSM (e.g., Cd$_3$As$_2$, Na$_3$Bi) as a function of rotating magnetic field -- we now use the following lattice model for a type-I DSM valid in the first Brillouin zone. The lattice model for type-I DSM can be written as,~\cite{Nagaosa_2014}

\begin{eqnarray}
H(\mathbf{k})&&=[M-t_{xy}(\cos k_x+\cos k_y)-t_z \cos k_z]\tau_{z}\sigma_{0}+ \nonumber \\
&& A_1 \sin k_x \tau_{x}\sigma_{z}- A_1 \sin k_y \tau_{y}\sigma_{0}+(\beta_1+\gamma_1)\sin k_z (\cos k_y \nonumber \\
&&-\cos k_x)\tau_{x}\sigma_{x}- (\beta_1-\gamma_1)\sin k_z \sin k_x \sin k_y\tau_{x}\sigma_{y}
\label{H_Lattice}
\end{eqnarray}
where $M$, $t_{xy}$, $t_z$, $A_1$, $\beta_1$ and $\gamma_1$ are real constants. As shown in Fig.~\ref{AH1} the behavior of AHC from the full lattice model given in Eq.~(\ref{H_Lattice}) is qualitatively consistent with the behavior of AHC from the low energy model given in Eq.~\ref{H_main}. In particular the anomalous Hall conductivity smoothly evolves as a function of polar angle $\theta$ near the mirror symmetric value $\theta=\pi/2$.

\section{Conclusion}
\label{summary}

In conclusion, quantum anomalies often manifest in observable consequences in condensed matter realizations of quantum systems such as Dirac and Weyl fermions. The most well known example is the chiral anomaly, which implies the non-conservation of chiral charge in Weyl semimetals in the presence of parallel electric and magnetic fields, resulting in observable signatures such as negative longitudinal magnetoresistance and planar Hall effect. Recently it has been argued that mirror symmetry in type-II DSMs across certain lattice planes constrains the behavior of the anomalous Hall effect with a rotating magnetic field. In particular, it has been shown that, due to an emergent mirror symmetry, the linearized theory of the type-II DSMs gives rise to a step-function dependence of the anomalous Hall conductivity on the sign of the departure of the polar angle ($\theta$) from the mirror symmetric value $\theta=\pi/2$ of the applied magnetic field.  This behavior, termed mirror anomaly in analogy with the similar behavior of anomalous Hall conductivity due to parity anomaly in 2D systems of massive Dirac fermions, has raised the interesting question if this behavior is generic for DSMs as negative longitudinal magnetoresistance and planar Hall effect. 

To answer this, we have computed the anomalous Hall conductivity for both type-I and type-II Dirac semimetals in the presence of an external magnetic field applied perpendicular to the mirror symmetric plane and rotated in the normal plane. 
Interestingly, we find that while the type-II DSMs possess mirror anomaly in AHC (defined as a step-function dependence of the AHC on the polar angle of the applied magnetic field applied perpendicular to the mirror symmetric plane), type-I DSMs show no such step function like behavior in AHC. Within the linearized theory of type-I DSM, the AHC vanishes even in the absence of time-reversal symmetry (applied magnetic field is along $z$ direction). Including cubic terms, we find that the AHC is non-zero in the presence of applied magnetic field, but evolves smoothly with the polar angle across the mirror symmetric value $\theta = \pi/2$, showing no evidence of mirror anomaly. In contrast to chiral anomaly, the observable consequences of mirror anomaly in DSMs are thus \textit{not} generic and can be used to distinguish between the types of DSMs in experiments. More generally, we have predicted the behavior of the anomalous Hall conductivity for type-I DSMs such as Cd$_{3}$As$_{2}$ and Na$_{3}$Bi which can be tested in experiments.

\section{Acknowledgement}
SN  acknowledges  MHRD, India for research fellowship. ST acknowledges support from ARO Grant No: (W911NF-16-1-0182).

\end{document}